\documentclass[aps,prb,twocolumn,shortbibliography,superscriptaddress,article]{revtex4-1}
\usepackage{epsfig}
\usepackage{epstopdf}
\usepackage{amsmath}
\usepackage{amsfonts}
\usepackage{amssymb}
\usepackage{hyperref}
\usepackage{bm}
\usepackage{makecell}
\usepackage{rotating}
\usepackage{hyperref}
\usepackage{multirow}
\usepackage{graphicx}
\usepackage{booktabs}
\usepackage{siunitx}
\usepackage{makecell}

\usepackage{graphicx}
\usepackage{dcolumn}
\usepackage{bm}
\usepackage{color}

\usepackage{tikz,xcolor,hyperref}

\definecolor{lime}{HTML}{A6CE39}
\DeclareRobustCommand{\orcidicon}{%
	\begin{tikzpicture}
	\draw[lime, fill=lime] (0,0)
	circle [radius=0.16]
	node[white] {{\fontfamily{qag}\selectfont \tiny ID}};
	\draw[white, fill=white] (-0.0625,0.095)
	circle [radius=0.007];
	\end{tikzpicture}
	\hspace{-2mm}
}

\foreach \x in {A, ..., Z}{%
	\expandafter\xdef\csname orcid\x\endcsname{\noexpand\href{https://orcid.org/\csname orcidauthor\x\endcsname}{\noexpand\orcidicon}}
}

\begin{document}

\title{Ferroelectric Altermagnetic Chern Insulator in magnetic field: \\ electrical control of the Chern number}

\author{Meysam Bagheri Tagani~\orcidA}
\email{mtagani@magtop.ifpan.edu.pl}
\affiliation{International Research Centre Magtop, Institute of Physics, Polish Academy of Sciences, Aleja Lotnik\'ow 32/46, PL-02668 Warsaw, Poland}
\affiliation{Department of Physics, University of Guilan, P. O. Box 41335-1914, Rasht, Iran}

\author{Carmine Autieri\orcidB}
\email{autieri@magtop.ifpan.edu.pl}
\affiliation{International Research Centre Magtop, Institute of Physics, Polish Academy of Sciences,
Aleja Lotnik\'ow 32/46, PL-02668 Warsaw, Poland}

\date{\today}
\begin{abstract}
We investigate electrically controllable Chern topology in a two-dimensional compensated (d)-wave altermagnet described by a lattice-regularized Bernevig--Hughes--Zhang model. In the altermagnetic reference state, the two Kramers sectors acquire momentum-dependent spin splitting and opposite sector Chern numbers, while the combined $C_{4z}\mathcal T$ symmetry enforces a vanishing total charge Chern number. We show that this hidden topological structure can be activated by orbital-selective magnetic coupling and independently tuned by ferroelectric orbital hybridization. The magnetic coupling removes the sector cancellation and generates $C=\pm1$ and $\pm2$ phases, whereas the polar distortion shifts the Dirac gap closing away from high-symmetry momenta and enables electrical transitions such as $C=0\rightarrow-1$ and $C=1\rightarrow2$. The resulting phases exhibit the expected chiral edge-state multiplicity and quantized anomalous Hall conductivity. Their topology is further reflected in the orbital magnetization through the in-gap relation $\partial_\mu \widetilde M_z=-C$. Finally, we show that the Chern phases remain robust against transverse Kramers-sector mixing and symmetry-allowed inversion-asymmetric spin--orbit coupling. These results establish a symmetry-based route to electrically tunable Chern insulating phases in compensated altermagnets.
\end{abstract}


\maketitle
	
\section{Introduction}
The realization of dissipationless edge transport remains a central objective in condensed matter physics, which can be useful for metrology applications\cite{10.1063/5.0233689}. Since the theoretical proposal and experimental observation of the quantum anomalous Hall effect (QAHE), Chern insulating phases have been predominantly associated with ferromagnets\cite{Wang_2015,Chang_2016,8vs2-jvc4,majewski2026valleystopologicalphases,PhysRevB.108.035121,tripathi2026universaltopologicalpowertransfer}. However, the QAHE in ferromagnets was observed only at very low temperatures so far\cite{Zhang_2016}.

The altermagnetism has emerged as a magnetic phase characterized by zero net magnetization in the non-relativistic limit and momentum-dependent spin splitting protected by rotational symmetry.\cite{doi:10.1126/sciadv.aaz8809,hayami2019momentum,Smejkal22beyond,autieri2312staggered,ssxp-gz9l,Fakhredine25b,Gonzalez2025-zc,Tenzin2025-ny,xt23-9pnv,leon2025strainenhancedaltermagnetismca3ru2o7} Spin-orbital effects as relativistic spin-momentum locking\cite{Fakhredine26} and orbital magnetization\cite{g32j-hnvz} has been shown to play an important role in altermagnets, where the relevance of the orbital angular momentum was supported by recent experimental observations\cite{ren2026atomicscaleobservationsymmetrybreaking}. Altermagnetism has stimulated intense interest in their transport and topological responses. In particular, altermagnets provide a promising route toward realizing Chern insulating phases without large macroscopic magnetization. Nevertheless, a fundamental obstacle persists. In altermagnets, the spin-up and spin-down states remain degenerate at the $\Gamma$ point in the nonrelativistic limit. This degeneracy obstructs the opening of a full topological gap required for the QAHE. Previous strategies to overcome this limitation relied on explicitly breaking rotational symmetry, thereby rendering magnetic sublattices inequivalent and driving the system into a ferrimagnetic phase. 
The rotational symmetry was broken by uniaxial strain\cite{Guo2023,10.1063/5.0147450,PhysRevB.107.214419}, by ferrimagnetism on the surface\cite{D3NR03681B,doi:10.1021/acs.nanolett.5c05341}, by stacking fault\cite{Li2025}, or by an external magnetic field\cite{doi:10.1021/acs.jpclett.6c01032,chen2026altermagnetsenablegateswitchablehelical}. 
In other materials with negligible magnetization, the Chern insulating phase with Chern number larger than 1 has been predicted in two-dimensional magnetic bilayers with mirror symmetry\cite{doi:10.1021/acs.nanolett.3c02489,doi:10.1021/acs.nanolett.6c00290} and in noncollinear magnets on a kagome lattice\cite{ahmed2025largechernnumberquantumanomalous}. The manipulation of the quantum anomalous Hall effect via ferroelectricity has been widely investigated primarily in MnBi$_2$Te$_4$, where time-reversal symmetry can be broken in different ways\cite{Li2025,q6zd-1z2r,doi:10.1021/acs.nanolett.5c00550,PhysRevB.110.205421,bai2026ferroelectricallyswitchablechiralitytopological}, whereas the corresponding phenomenon in altermagnets has received considerably less attention.

Among other topological effects in altermagnets or in system based on altermagnets, we can mention the quantum spin Hall\cite{PhysRevLett.134.096703,s57q-q7gt,doi:10.1021/acs.jpclett.6c01032, tagani2026berry}, the quantization of the spin circular photogalvanic effect \cite{463y-q7lt}, the topological piezomagnetic effect\cite{radhakrishnan2026topologicalpiezomagneticeffecttwodimensional}, the layer Hall effect\cite{vwtc-klg7}, axion insulator\cite{PhysRevB.103.195308}, almost half-quantized planar Hall effects\cite{zt4l-y18j,1b5p-k5vj}, and higher-order topology with corner states\cite{9wcm-pmr2,kplp-819f}.

In this work, we demonstrate that the interplay among the external magnetic field, ferroelectricity, and spin canting can lift the $\Gamma$-point degeneracy, yielding a rich phase diagram with Chern insulating phases. Specifically, we construct and analyze a two-dimensional $d$-wave altermagnetic model coupled to a ferroelectric order parameter. The resulting phase is a ferroelectric Chern insulator that exhibits a fully gapped bulk spectrum and quantized Hall conductivity in the absence of any external magnetic field. Crucially, the ferroelectric polarization acts as an electrically tunable control parameter between different topological phases $C=\pm1$ and $C=\pm2$; the system therefore realizes an electrically switchable topological phase based on altermagnets.

\section{Model and symmetry analysis}
\label{sec:model}

We consider a minimal four-band model on a two-dimensional square lattice that combines band inversion, compensated $d$-wave altermagnetism, ferroelectric orbital hybridization, and magnetic-field-induced spin mixing.  The basis is
\begin{equation}
\Psi_{\mathbf k}=
\left(|A,+\rangle,|A,-\rangle,|B,+\rangle,|B,-\rangle\right)^T,
\label{eq:basis}
\end{equation}
where $A$ and $B$ are opposite-parity orbital doublets with $|j_z|=1/2$ and $|j_z|=3/2$, respectively, and $\pm$ label the Kramers partners of the nonmagnetic parent.  The Pauli matrices $\tau_i$ act in the $A/B$ orbital space and $s_i$ in the Kramers-sector space.  When spin is approximately conserved, $s_z$ tracks the corresponding physical-spin polarization.  Tensor products are implicit below.

Time reversal and inversion are represented by
\begin{equation}
\mathcal T=\tau_0 i s_y K,
\qquad
\mathcal P=\tau_z s_0,
\label{eq:TP}
\end{equation}
with $K$ complex conjugation.  We adopt the clockwise fourfold rotation $C_{4z}:(k_x,k_y)\mapsto(k_y,-k_x)$, represented in the basis of Eq.~(\ref{eq:basis}) by
\begin{align}
U_{C_{4z}}&=
\exp\!\left[-\frac{i\pi}{2}\left(\tau_0-\frac{\tau_z}{2}\right)s_z\right] \\ \nonumber
&=\operatorname{diag}\!\left(e^{-i\pi/4},e^{i\pi/4},e^{-3i\pi/4},e^{3i\pi/4}\right).
\label{eq:C4operator}
\end{align}
It satisfies $U_{C_{4z}}^4=- 1$ and, in particular,
\begin{align}
U_{C_{4z}}(\tau_x s_z)U_{C_{4z}}^\dagger &=-\tau_y s_0, \\ \nonumber
U_{C_{4z}}\tau_ys_0 U_{C_{4z}}^\dagger &=\tau_x s_z,\\ \nonumber
U_{C_{4z}}\tau_zs_0 U_{C_{4z}}^\dagger &=\tau_zs_0.
\label{eq:C4transformations}
\end{align}

The working Hamiltonian is
\begin{equation}
H(\mathbf k)=H_{\rm BHZ}(\mathbf k)+H_{\rm AM}(\mathbf k)+H_{\mathcal B}+H_{\rm FE}+H_{\rm SOC}(\mathbf k),
\label{eq:Hfull}
\end{equation}
where the five terms describe, respectively, the band-inverted parent, compensated altermagnetic exchange, magnetic-field coupling, ferroelectric orbital hybridization, and inversion-asymmetric spin--orbit coupling.

\subsection{Band-inverted parent}

The parent is the lattice-regularized BHZ Hamiltonian~\cite{bernevig2006quantum,https://doi.org/10.1002/advs.202522203}
\begin{equation}
H_{\rm BHZ}(\mathbf k)=
M_{\mathbf k}\tau_zs_0
+A\sin k_x\,\tau_x s_z
+A\sin k_y\,\tau_ys_0,
\label{eq:HBHZ}
\end{equation}
where $M_{\mathbf k}=M_0-2B_0\left(2-\cos k_x-\cos k_y\right)$. Here $M_0$ is the band-inversion mass, $B_0$ controls the lattice curvature, and $A$ sets the interorbital Dirac velocity.  The Hamiltonian is invariant under $\mathcal T$, $\mathcal P$, and $C_{4z}$ as defined above.

Because $[H_{\rm BHZ},s_z]=0$, the two Kramers sectors decouple.  In the reordered basis $(|A,+\rangle,|B,+\rangle,|A,-\rangle,|B,-\rangle)^T$, we have
\begin{equation}
h_{\pm}=M_{\mathbf k}\tau_z\pm A\sin k_x\,\tau_x+A\sin k_y\,\tau_y,
\qquad
h_-(\mathbf k)=h_+^*(-\mathbf k).
\label{eq:BHZblocks}
\end{equation}
The two blocks therefore carry opposite Chern numbers, $C_+=-C_-$, and the charge Chern number of the time-reversal-invariant parent vanishes.

\subsection{Compensated altermagnetic order}

We model the $d$-wave altermagnetic exchange by
\begin{equation}
H_{\rm AM}(\mathbf k)=f_d(\mathbf k)\left(\Delta_0\tau_0+\Delta_z\tau_z\right)s_z,
\label{eq:HAM}
\end{equation}
wher $f_d(\mathbf k)=\cos k_x-\cos k_y$. The coefficient $\Delta_0$ gives the leading momentum-dependent spin splitting, while $\Delta_z$ describes an orbital-dependent component that enters the Dirac mass and allows the two Kramers sectors to undergo distinct inversions.  Since $f_d(k_y,-k_x)=-f_d(k_x,k_y)$ and $s_z$ is odd under time reversal, $H_{\rm AM}$ breaks $\mathcal T$ and $C_{4z}$ separately but preserves their product $\mathcal S=C_{4z}\mathcal T$.  The splitting vanishes where $f_d=0$, including the $\Gamma$ and $M$ points and the Brillouin-zone diagonals.

The compensated reference Hamiltonian $H_{\rm ref}=H_{\rm BHZ}+H_{\rm AM}$ therefore obeys $\mathcal S H_{\rm ref}(\mathbf k)\mathcal S^{-1}=H_{\rm ref}(\mathcal S\mathbf k)$.  Its Berry curvature satisfies
\begin{equation}
\Omega_z(\mathbf k)=-\Omega_z(\mathcal S\mathbf k),
\qquad
C=\frac{1}{2\pi}\int_{\rm BZ}d^2k\,\Omega_z(\mathbf k)=0.
\label{eq:Czero}
\end{equation}
Thus altermagnetism produces momentum-dependent spin splitting and nontrivial local Berry curvature, while $C_{4z}\mathcal T$ enforces a vanishing total charge Chern number in the compensated reference state.

\subsection{Magnetic-field coupling and transverse canting}

We retain the Zeeman/exchange part of the external magnetic field and neglect orbital Landau quantization.  The two field-induced couplings used in the calculations are
\begin{equation}
H_{\mathcal B}=b_o\,\tau_z s_z+b_x\,\tau_0 s_x.
\label{eq:HB}
\end{equation}
Here $b_o$ is the orbital-difference longitudinal Zeeman/exchange energy and is taken proportional to the out-of-plane field component, whereas $b_x$ is a uniform transverse spin-mixing energy associated with a tilted field or field-induced canting.  The first term changes the Dirac masses of the two Kramers sectors in opposite directions and is the field parameter that drives the Chern transitions discussed below.  Its matrix structure is identical to a momentum-independent $\tau_zs_z$ exchange field, but in the present model its physical origin is externally controlled and is distinct from the intrinsic $d$-wave altermagnetic mass $\Delta_z f_d(\mathbf k)\tau_zs_z$.  The transverse term mixes the two Kramers sectors and removes exact $s_z$ conservation.

An orbital-average longitudinal Zeeman term $b_u\tau_0s_z$ is symmetry allowed.  It shifts the two Kramers sectors without changing their eigenvectors or direct Chern boundaries and is set to zero in all numerical calculations; we therefore omit it from the working Hamiltonian.  Both terms retained in Eq.~(\ref{eq:HB}) break $C_{4z}\mathcal T$ for a fixed nonzero field configuration.

\subsection{Ferroelectric orbital hybridization}

We parameterize the electronic coupling to the in-plane polarization $\mathbf P=(P_x,P_y)$ by $p_i=v_P P_i$.  The minimal inversion-odd orbital hybridization consistent with the double-group representation is
\begin{equation}
H_{\rm FE}=p_x\,\tau_x+p_y\,\tau_y s_z.
\label{eq:HFEvector}
\end{equation}
The relative sign is fixed by the clockwise convention: $U_{C_{4z}}\tau_xU_{C_{4z}}^\dagger=-\tau_y s_z$ and $U_{C_{4z}}(\tau_y s_z)U_{C_{4z}}^\dagger=\tau_x$, so that $U_{C_{4z}}H_{\rm FE}(\mathbf P)U_{C_{4z}}^\dagger=H_{\rm FE}(C_{4z}\mathbf P)$ with $C_{4z}(P_x,P_y)=(P_y,-P_x)$.  Inversion gives $\mathcal P H_{\rm FE}(\mathbf P)\mathcal P^{-1}=H_{\rm FE}(-\mathbf P)$.

All calculations use a domain polarized along $x$, for which $p_y=0$ and
\begin{equation}
H_{\rm FE}=p\,\tau_x,
\qquad p\equiv p_x=v_P P_x.
\label{eq:HFE}
\end{equation}
A fixed $p\neq0$ selects a crystallographic direction, breaks $C_{4z}$ and therefore $C_{4z}\mathcal T$, and removes the Berry-curvature cancellation of the compensated altermagnet.  Since time reversal is already broken by $H_{\rm AM}$, this symmetry lowering can activate a nonzero Chern number without a uniform ferromagnetic moment.

\subsection{Inversion-asymmetric spin--orbit coupling}

The conventional orbital-independent Rashba form $\sin k_y s_x-\sin k_xs_y$ is not covariant under the double-group $C_{4z}$ representation above.  Indeed,
\begin{align}
U_{C_{4z}}\tau_0s_xU_{C_{4z}}^\dagger &=\tau_zs_y,
\\ \nonumber
U_{C_{4z}}\tau_0s_yU_{C_{4z}}^\dagger &=-\tau_zs_x,
\\ \nonumber
U_{C_{4z}}\tau_zs_xU_{C_{4z}}^\dagger &=\tau_0s_y,
\\ \nonumber
U_{C_{4z}}\tau_zs_yU_{C_{4z}}^\dagger &=-\tau_0s_x. \nonumber
\label{eq:C4_SOC_matrices}
\end{align}
The leading linear-in-momentum spin-mixing structures used in the robustness analysis are therefore
\begin{equation}
\begin{aligned}
H_{R1}(\mathbf k)&=\lambda_1\left(\sin k_y\,\tau_0s_x+\sin k_x\,\tau_zs_y\right),\\
H_{R2}(\mathbf k)&=\lambda_2\left(\sin k_y\,\tau_zs_x+\sin k_x\,\tau_0s_y\right),
\end{aligned}
\label{eq:HR12}
\end{equation}
with $H_{\rm SOC}=H_{R1}+H_{R2}$. Both terms are time-reversal even and $C_{4z}$ covariant.  For a fixed nonzero $\lambda_{1,2}$, they break inversion; if the couplings originate from an inversion-odd structural asymmetry, inversion covariance is recovered by reversing $\lambda_{1,2}$ together with $\mathbf k$.  We therefore treat $\lambda_1$ and $\lambda_2$ as independent phenomenological SOC energies that probe the robustness of the ferroelectric Chern phases against generic spin-sector mixing.

For completeness, the in-plane polar order also allows SOC that is odd in momentum but diagonal in $s_z$.  A representative $C_{4z}$-compatible channel is
\begin{equation}
H_{P{\rm SO}}(\mathbf k)=\left(p_x\sin k_y-p_y\sin k_x\right)\left(\eta_0\tau_0+\eta_z\tau_z\right)s_z.
\label{eq:HPSO}
\end{equation}
The coefficients $\eta_0$ and $\eta_z$ are not additional parameters in the numerical phase diagrams; this term is used only to discuss the symmetry of the polar SOC channel.  For the $x$-polarized domain, $H_{P{\rm SO}}\propto p\sin k_y$ and therefore vanishes at the ferroelectric Dirac closings, where $\sin k_y=0$.  It can nevertheless modify the indirect gap and Berry-curvature distribution away from the critical momentum.  Without specifying additional vertical mirrors, Eq.~(\ref{eq:HPSO}) should be viewed as a representative rather than unique polar SOC invariant.

The resulting hierarchy is deliberate.  At $p=b_o=b_x=\lambda_1=\lambda_2=0$, the compensated altermagnetic reference preserves $C_{4z}\mathcal T$ and has $C=0$.  Ferroelectricity and the longitudinal orbital-dependent field provide two distinct routes for lifting this cancellation, while $b_x$ and $H_{\rm SOC}$ test the stability of the resulting Chern phases once the spin sectors are mixed.  This separation allows the origin of the topology to be distinguished from relativistic corrections and canting effects.

                                   
\section{Results}
\label{sec:results}

\subsection{Compensated altermagnetic parent}
\label{subsec:analytic_parent}

We begin with the spin-conserving limit, which provides the reference point for the field- and ferroelectricity-driven transitions discussed below.  We set $b_x=\lambda_1=\lambda_2=0$ and take the polar distortion along $x$, $p\equiv v_P P_x$ with $P_y=0$.  Since $[H(\mathbf k),s_z]=0$, the Hamiltonian separates into two sectors, $h_s(\mathbf k)=\varepsilon_s(\mathbf k)\tau_0+\mathbf d_s(\mathbf k)\cdot\boldsymbol\tau$, with $s=\pm1$, as
\begin{equation}
  \qquad
 \begin{aligned}
 \varepsilon_s&=s\left(\Delta_0f_{\mathbf k}+b_u\right),\\
 d_{x,s}&=sA\sin k_x+p,\qquad d_{y,s}=A\sin k_y,\\
 d_{z,s}&=M_{\mathbf k}+s\left(\Delta_zf_{\mathbf k}+b_o\right),
 \end{aligned}
 \label{eq:block_hamiltonian}
\end{equation}
with eigenvalues $E_{s,\eta}=\varepsilon_s+\eta|\mathbf d_s|$ ($\eta=\pm1$).  This form separates the energetic and topological roles of the couplings: $\Delta_0f_{\mathbf k}$ and $b_u$ multiply $\tau_0$ and therefore shift the two sectors without changing their Bloch eigenvectors, whereas $\Delta_z$, $b_o$, and $p$ modify $\mathbf d_s$ and hence the Berry curvature and topological phase boundaries.  The former terms can nevertheless close the global indirect gap by producing an energetic overlap between the sectors.

\begin{figure}[t]
 \centering
 \includegraphics[width=\linewidth]{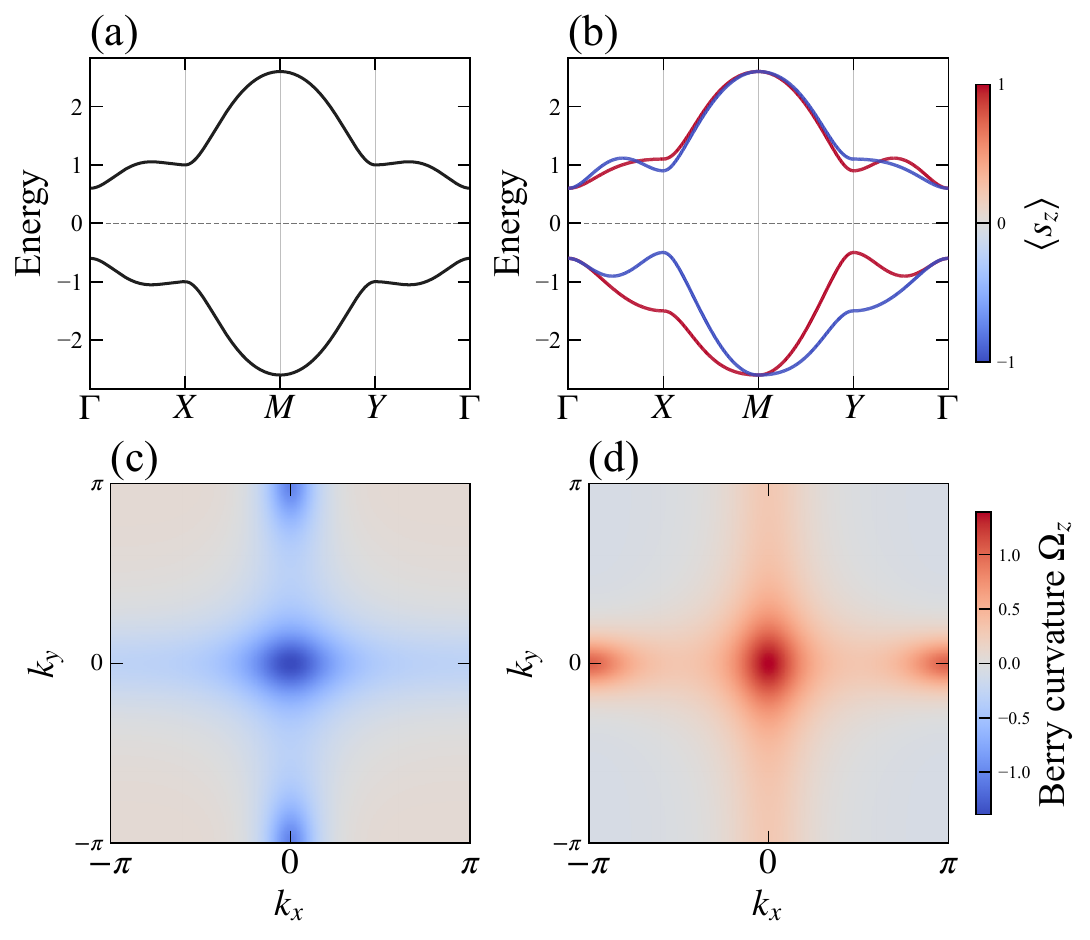}
 \caption{\label{fig:reference_bands_berry}
 (a) Band structure of the time-reversal-invariant BHZ parent.  (b) Band structure after including the $d$-wave altermagnetic couplings $\Delta_0$ and $\Delta_z$; color denotes $\langle s_z\rangle$.  The splitting vanishes at $\Gamma$ and $M$ and reverses under $C_{4z}$.  (c),(d) Berry curvature of the occupied bands in the $s=+1$ and $s=-1$ sectors, carrying $C_+=-1$ and $C_-=+1$, respectively.  Parameters are $A=1$, $B_0=0.40$, $M_0=0.60$, $\Delta_0=0.10$, and $\Delta_z=0.15$; all other couplings vanish.}
\end{figure}

For the compensated altermagnetic reference state, $p=b_u=b_o=0$.  The $d$-wave form factor satisfies $f_{R_{4z}\mathbf k}=-f_{\mathbf k}$ for the clockwise rotation $R_{4z}(k_x,k_y)=(k_y,-k_x)$. Consequently, the spin splitting changes sign under $C_{4z}$ and vanishes on $f_{\mathbf k}=0$, including the $\Gamma$ and $M$ points and the Brillouin-zone diagonals.  This behavior is visible in Fig.~\ref{fig:reference_bands_berry}(b) and distinguishes the altermagnetic splitting from a uniform Zeeman shift.

For each two-band sector, the Berry curvature is
\begin{equation}
 \Omega_{s,\eta}(\mathbf k)=
 \frac{\eta}{2}
 \frac{\mathbf d_s\cdot\left(\partial_{k_x}\mathbf d_s\times\partial_{k_y}\mathbf d_s\right)}{|\mathbf d_s|^3}.
 \label{eq:block_Berry}
\end{equation}
The combined antiunitary symmetry $\mathcal S=C_{4z}\mathcal T$ relates the two sectors according to
\begin{equation}
 \Omega_{s,\eta}(k_x,k_y)=-\Omega_{-s,\eta}(-k_y,k_x),
 \qquad C_s=-C_{-s}.
 \label{eq:C4T_Berry_relation}
\end{equation}
Thus the compensated state may carry strongly structured sector-resolved Berry curvature while its total charge Chern number vanishes exactly.  This cancellation is the topological counterpart of the $C_{4z}\mathcal T$ constraint on the linear anomalous Hall response in a pure $d$-wave altermagnet~\cite{takahashi2025elasto,mukherjee2026second}.

At $p=0$, gap closings occur only at the four time-reversal-invariant momenta.  The corresponding Dirac masses are
\begin{equation}
 \begin{aligned}
 m_s^{\Gamma}&=M_0, & m_s^{M}&=M_0-8B_0,\\
 m_s^{X}&=M_0-4B_0-2s\Delta_z, &
 m_s^{Y}&=M_0-4B_0+2s\Delta_z,
 \end{aligned}
 \label{eq:reference_TRIM_masses}
\end{equation}
and the occupied band of sector $s$ has
\begin{equation}
 C_s=-\frac{s}{2}\left[\operatorname{sgn}m_s^{\Gamma}+\operatorname{sgn}m_s^{M}-\operatorname{sgn}m_s^{X}-\operatorname{sgn}m_s^{Y}\right].
 \label{eq:reference_block_chern}
\end{equation}
The interchange $m_{-s}^{X}=m_s^{Y}$ and $m_{-s}^{Y}=m_s^{X}$ immediately gives $C_{-s}=-C_s$.  A finite orbital-selective altermagnetic coupling therefore separates the $X$- and $Y$-point inversions without generating a net charge Chern number.

Figure~\ref{fig:reference_bands_berry} illustrates this reference state for $A=1$, $B_0=0.40$, $M_0=0.60$, $\Delta_0=0.10$, and $\Delta_z=0.15$.  The BHZ parent is doubly degenerate, whereas the altermagnetic terms generate the expected momentum-dependent splitting while preserving the $\Gamma$- and $M$-point degeneracies.  Equation~(\ref{eq:reference_block_chern}) gives $C_+=-1$ and $C_-=+1$, while a non-Abelian Fukui calculation for the two occupied bands gives $C=0$~\cite{fukui2005chern}.  The direct and indirect gaps are $1.20$ and $1.10$, respectively.  The two Berry-curvature distributions in Figs.~\ref{fig:reference_bands_berry}(c) and \ref{fig:reference_bands_berry}(d) integrate to opposite integers and satisfy Eq.~(\ref{eq:C4T_Berry_relation}) numerically.

\begin{figure*}[t]
\centering
\includegraphics[width=0.70\textwidth]{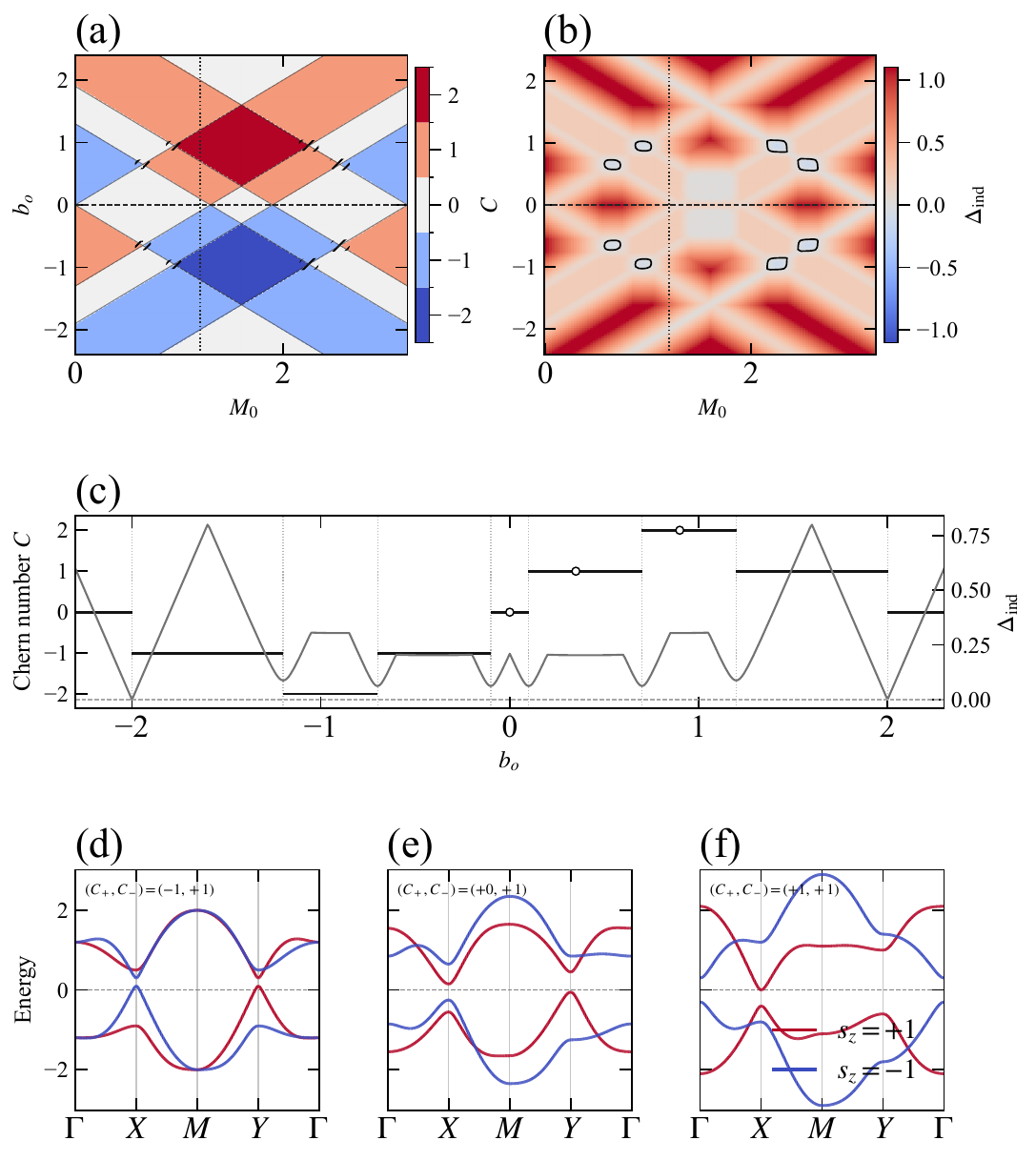}
\caption{\label{fig:field_topology}
Magnetic-field-induced Chern topology.  (a) Exact $C(M_0,b_o)$ for $A=1$, $B_0=0.40$, $\Delta_0=0.10$, $\Delta_z=0.15$, and $b_u=0$; thin lines mark the analytical conditions $m_s^{\Gamma,X,Y,M}=0$, and hatching denotes a negative indirect gap.  (b) Indirect gap on the same parameter plane; the black contour marks $\Delta_{\rm ind}=0$.  (c) Field cut at $M_0=1.20$, showing the sequence $C=0\to1\to2\to1\to0$.  (d)--(f) Band structures along $\Gamma$--$X$--$M$--$Y$--$\Gamma$ for representative $C=0$, $1$, and $2$ phases; color denotes $\langle s_z\rangle$.}
\end{figure*}

\subsection{Magnetic-field-induced Chern topology}
\label{subsec:field_topology}

We next isolate the longitudinal magnetic response by setting $p=b_x=\lambda_1=\lambda_2=0$.  The block form in Eq.~(\ref{eq:block_hamiltonian}) remains valid.  The uniform term $b_u\tau_0s_z$ shifts the two sectors in energy, whereas the orbital-dependent term $b_o\tau_zs_z$ enters the Dirac mass and changes the band inversion independently in the two sectors.  The four masses become

 \begin{align}
 \label{eq:field_TRIM_masses}
 m_s^{\Gamma}&=M_0+s b_o, \\ \nonumber
  m_s^{M}&=M_0-8B_0+s b_o,\\ \nonumber
 m_s^{X}&=M_0-4B_0+s(b_o-2\Delta_z),\\ \nonumber
 m_s^{Y}&=M_0-4B_0+s(b_o+2\Delta_z).
 \end{align}

Together with Eq.~(\ref{eq:reference_block_chern}), these masses give the exact phase boundaries shown in Fig.~\ref{fig:field_topology}(a).  A finite $b_o$ removes the $C_{4z}\mathcal T$-enforced cancellation and allows one sector to change topology while the other remains unchanged, naturally producing odd-Chern phases.  Field reversal exchanges the two sectors and reverses the charge Chern number, $C(-b_o)=-C(b_o)$, as long as the insulating gap remains open.

A nonzero Chern index corresponds to a Chern insulator only when the occupied and unoccupied manifolds are globally separated.  We therefore monitor the half-filled indirect gap
\begin{equation}
 \Delta_{\rm ind}=\min_{\mathbf k}E_3(\mathbf k)-\max_{\mathbf k}E_2(\mathbf k),
 \qquad E_1\le E_2\le E_3\le E_4.
 \label{eq:indirect_gap}
\end{equation}
Figure~\ref{fig:field_topology}(b) shows that broad portions of the $|C|=1$ and $|C|=2$ sectors remain fully insulating.  The distinction is particularly relevant for $\Delta_0$ and $b_u$, which leave the sector Chern numbers unchanged but can modify $\Delta_{\rm ind}$.

For the representative cut $M_0=1.20$, with $A=1$, $B_0=0.40$, $\Delta_0=0.10$, and $\Delta_z=0.15$, the positive-field transitions occur at $b_o=0.10$, $0.70$, $1.20$, and $2.00$.  The corresponding sequence is
\begin{equation}
 C=0\ \longrightarrow\ 1\ \longrightarrow\ 2\ \longrightarrow\ 1\ \longrightarrow\ 0,
 \label{eq:field_sequence}
\end{equation}
as shown in Fig.~\ref{fig:field_topology}(c).  The first two transitions successively change the $s=+$ sector from $-1$ to $0$ and then to $+1$; the latter two remove the $s=-$ and $s=+$ topological sectors.  The band structures in Figs.~\ref{fig:field_topology}(d)--\ref{fig:field_topology}(f) illustrate representative $C=0$, $1$, and $2$ phases.  Independent block and non-Abelian Fukui calculations reproduce the analytical integers.

\begin{figure*}[t]
    \centering
    \includegraphics[width=0.80\textwidth]{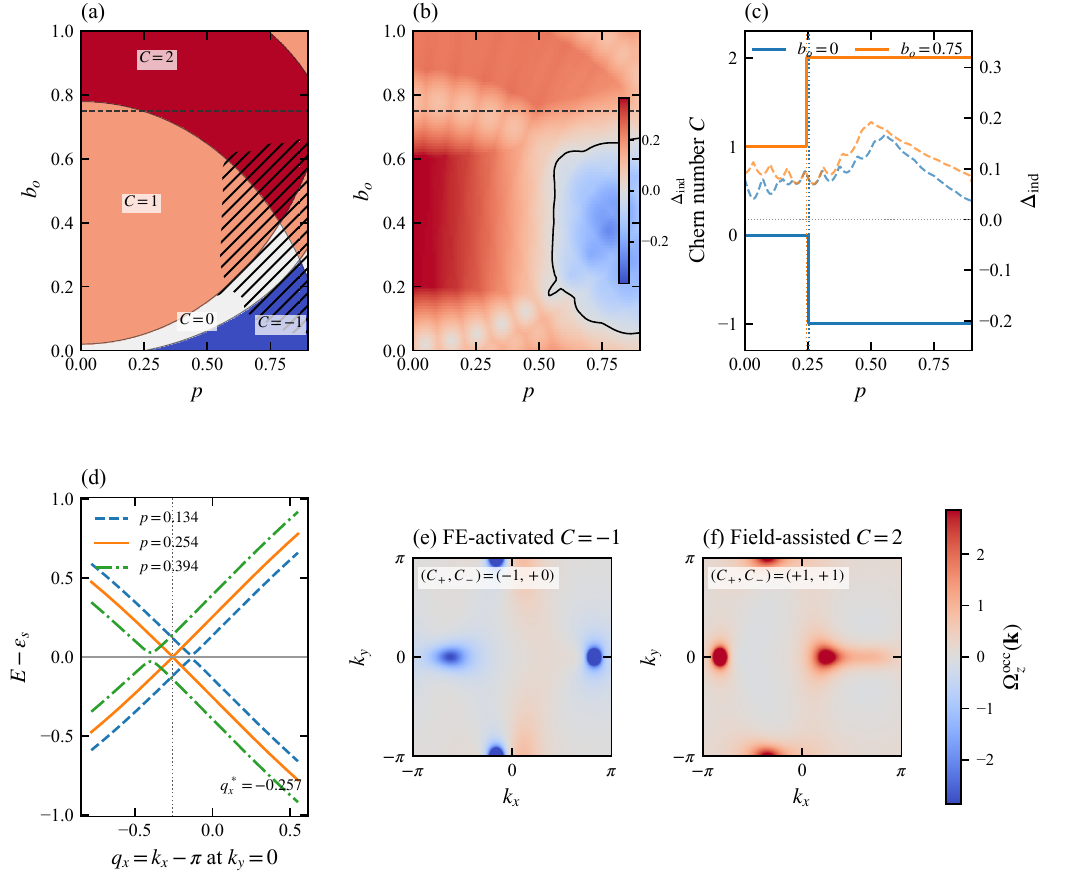}
    \caption{\label{fig:FEcontrol}
    Ferroelectric activation and field-assisted control of Chern topology.  (a) Exact $C(p,b_o)$; thin curves mark $m_{s,\eta\nu}=0$, and hatched regions are metallic at half filling.  (b) Indirect gap on the same plane.  (c) Polarization cuts at $b_o=0$ and $0.75$, showing $C=0\to-1$ and $C=1\to2$, respectively.  (d) Centered spectrum of the critical $s=-1$ block near $X$, showing the finite-momentum closing at $q_x^*\simeq-0.257$.  (e) Occupied Berry curvature for the zero-field $C=-1$ phase at $p=0.55$.  (f) Occupied Berry curvature for the field-assisted $C=2$ phase at $p=0.55$ and $b_o=0.75$.  Parameters are $A=1$, $B_0=0.40$, $M_0=1.20$, $\Delta_0=0.10$, and $\Delta_z=0.19$.}
\end{figure*}

\subsection{Ferroelectric activation and electrical control}
\label{subsec:FEtopology}

We now restore the $x$-polarized hybridization $p=v_P P_x$ while retaining the spin-conserving limit.  Ferroelectricity enters Eq.~(\ref{eq:block_hamiltonian}) through $d_{x,s}=sA\sin k_x+p$ and therefore shifts the kinetic zero rather than the Dirac mass itself.  A gap closing satisfies
\begin{equation}
 \cos k_x^*=\eta c_p,\quad \cos k_y^*=\nu,
 \label{eq:FEkineticzeros}
\end{equation}
with $\eta,\nu=\pm1$, $|p|<|A|$, and $c_p\equiv\sqrt{1-\frac{p^2}{A^2}}$.  The mass at each displaced Dirac point is
\begin{equation}
 m_{s,\eta\nu}=M_0-4B_0+s b_o+\nu(2B_0-s\Delta_z)+\eta c_p(2B_0+s\Delta_z).
 \label{eq:FEmass}
\end{equation}
Hence $m_{s,\eta\nu}=0$ gives the phase boundaries analytically.  Equivalently,
\begin{align}
 c_c^{s\eta\nu}&=-\eta\frac{M_0-4B_0+s b_o+\nu(2B_0-s\Delta_z)}{2B_0+s\Delta_z},
 \\ \nonumber
 |p_c|&=|A|\sqrt{1-(c_c^{s\eta\nu})^2},
 \label{eq:FEcriticalp}
\end{align}
provided $0\le c_c^{s\eta\nu}\le1$.  The chirality of the corresponding Dirac point is $s\eta\nu$, yielding
\begin{equation}
 C_s=-\frac12\sum_{\eta,\nu=\pm1}s\eta\nu\,\operatorname{sgn}m_{s,\eta\nu},
 \qquad C=C_++C_-.
 \label{eq:FEChern}
\end{equation}
Equations~(\ref{eq:FEmass})--(\ref{eq:FEChern}) therefore provide a closed analytical classification of the spin-conserving ferroelectric phase diagram.

Figure~\ref{fig:FEcontrol}(a) shows the resulting $C(p,b_o)$ map for $A=1$, $B_0=0.40$, $M_0=1.20$, $\Delta_0=0.10$, and $\Delta_z=0.19$.  At $b_o=0$, the nonpolar state has $(C_+,C_-)=(-1,+1)$ and $C=0$.  The first ferroelectric transition occurs at $p_c\simeq0.254$ in the $s=-1$ sector, at the finite momentum $(k_x^*,k_y^*)\simeq(2.88,0)$.  Across the transition $C_-$ changes from $+1$ to $0$, producing a zero-field $C=-1$ phase.  At $p=0.55$ this state remains insulating with $\Delta_{\rm ind}\simeq0.15$.

The longitudinal field provides a second control axis.  At $b_o=0.75$, the nonpolar state has $C=1$, while increasing $p$ drives the $s=+1$ sector through a displaced Dirac closing at $p_c\simeq0.244$ and produces $C=2$.  At $p=0.55$, the indirect gap is approximately $0.17$.  Ferroelectricity therefore changes the number of chiral channels by selectively driving one altermagnetic sector through a finite-momentum inversion.

The closing and reopening are resolved directly in Fig.~\ref{fig:FEcontrol}(d), where the critical momentum lies at $q_x^*=k_x^*-\pi\simeq-0.257$.  Because the transition occurs away from a high-symmetry point, it would be missed by inspecting only the conventional $\Gamma$--$X$--$M$--$Y$--$\Gamma$ path.  The Berry-curvature maps in Figs.~\ref{fig:FEcontrol}(e) and \ref{fig:FEcontrol}(f) show the corresponding redistribution: the zero-field polar state carries a net flux $C=-1$, whereas in the field-assisted phase the two contributions add to $C=2$.

A symmetry constraint of the minimal model is also worth noting.  Since the phase boundaries depend on $p$ only through $c_p=\sqrt{1-p^2/A^2}$,
\begin{equation}
 C(+p,b_o)=C(-p,b_o).
 \label{eq:PevenChern}
\end{equation}
Thus, the inversion-related $\pm P_x$ domains carry the same Chern number in the present Hamiltonian.  Electrical control refers here to tuning the magnitude of the polar distortion, or switching between nonpolar and polar states; reversing the Hall chirality by $P_x\to-P_x$ would require an additional symmetry-breaking ingredient that makes the two domains topologically inequivalent.

\subsection{Bulk--boundary correspondence and Hall response}
\label{subsec:bulk_boundary}

The bulk invariants can be tested independently through edge-state spectral flow and the anomalous Hall conductivity.  We construct a ribbon periodic along $x$ and open along $y$ by Fourier transforming the lattice Hamiltonian in $k_y$; no boundary-specific terms are introduced.  Edge localization is identified from the probability weight in the outermost unit cells of each boundary.

\begin{figure*}[t]
\centering
\includegraphics[width=0.70\textwidth]{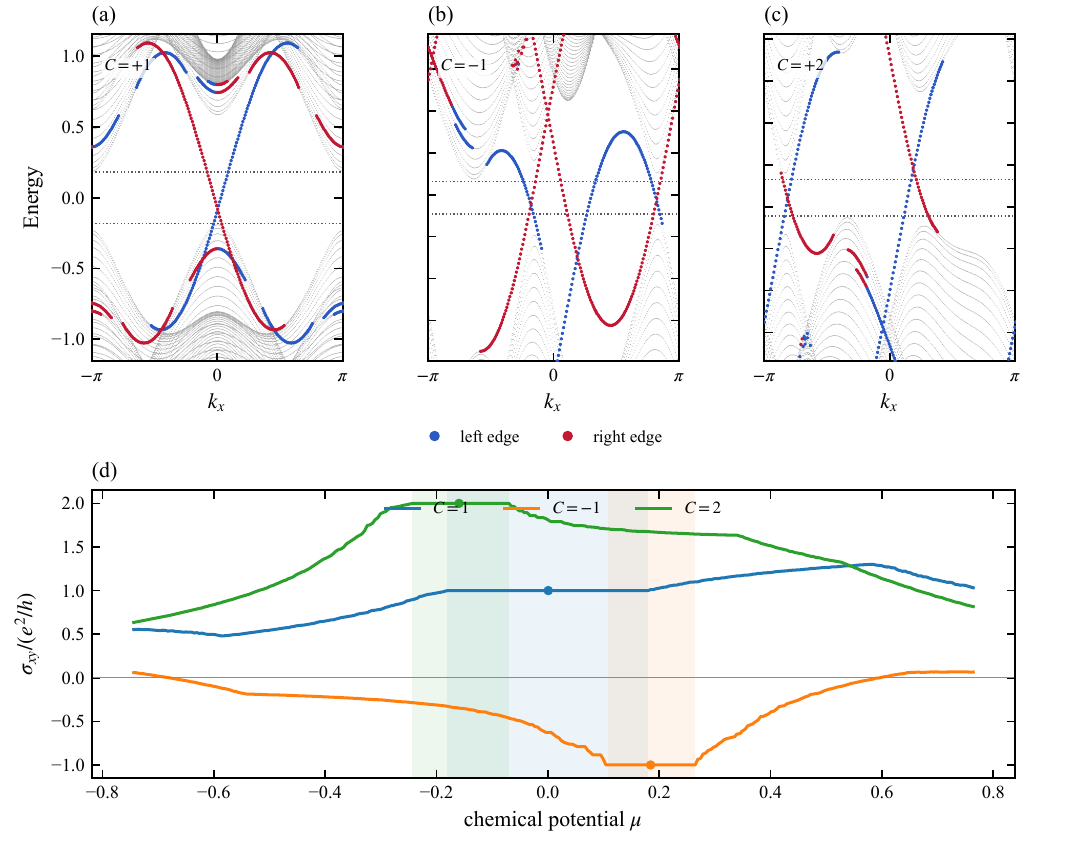}
\caption{\label{fig:bulk_boundary}
Bulk--boundary correspondence and quantized anomalous Hall transport.  (a)--(c) Ribbon spectra for representative $C=1$, $C=-1$, and $C=2$ phases.  Gray states denote the projected bulk continuum; blue and red markers denote states localized at opposite edges.  The $C=2$ phase hosts two co-propagating channels per edge.  (d) Hall conductivity versus chemical potential for the same phases; shaded intervals mark the global bulk gaps, within which $\sigma_{xy}=C e^2/h$.  Common parameters are $A=1$, $B_0=0.40$, $M_0=1.20$, $\Delta_0=0.10$, and $\Delta_z=0.19$.}
\end{figure*}

Figures~\ref{fig:bulk_boundary}(a)--\ref{fig:bulk_boundary}(c) show representative ribbon spectra.  The field-induced state at $p=0$ and $b_o=0.40$ has $C=1$ and one chiral branch per edge.  The zero-field ferroelectric state at $p=0.55$ has $C=-1$ and the opposite spectral flow.  For $p=0.55$ and $b_o=0.75$, two co-propagating branches appear on each edge, as required for $C=2$.  One crossing is continued through the periodic boundary $k_x=\pm\pi$, but the net edge winding is two, in agreement with bulk--boundary correspondence~\cite{hatsugai1993chern}.

The Hall conductivity is evaluated from the full four-band Berry curvature using the standard Kubo expression,
\begin{equation}
 \sigma_{xy}(\mu)=\frac{e^2}{\hbar}\sum_n\int_{\rm BZ}\frac{d^2k}{(2\pi)^2}
 \Theta(\mu-E_{n\mathbf k})\,\Omega_{n,z}(\mathbf k).
 \label{eq:hall_mu}
\end{equation}
Inside a global bulk gap, this reduces to the TKNN result $\sigma_{xy}=C e^2/h$~\cite{thouless1982quantized}.  Figure~\ref{fig:bulk_boundary}(d) shows plateaus at $+1$, $-1$, and $+2$ in units of $e^2/h$ for the same three phases.  Their indirect gaps are approximately $0.36$, $0.16$, and $0.17$, respectively.  The agreement among the bulk Chern number, edge-channel multiplicity, and Hall plateau confirms that these states are genuine Chern insulators rather than locally inverted or metallic phases~\cite{hatsugai1993chern,thouless1982quantized,liu2008quantum}.

\subsection{Orbital magnetization}
\label{subsec:orbitalmag}

Orbital magnetization provides a complementary thermodynamic probe of the same topology.  We evaluate it using the modern bulk theory of orbital magnetism~\cite{thonhauser2005orbital,ceresoli2006orbital,bianco2013orbital}.  At zero temperature,
\begin{equation}
 \begin{aligned}
 M_z^{\rm orb}={}&\frac{e}{\hbar}\sum_n\int_{\rm BZ}\frac{d^2k}{(2\pi)^2}
 \Theta(\mu-E_{n\mathbf k})\\
 &\times\operatorname{Im}\left\langle\partial_{k_x}u_{n\mathbf k}\left|
 H_{\mathbf k}+E_{n\mathbf k}-2\mu
 \right|\partial_{k_y}u_{n\mathbf k}\right\rangle .
 \end{aligned}
 \label{eq:modern_Morb}
\end{equation}
with $e>0$. Numerically, we use the equivalent sum-over-states form, which avoids derivatives of the eigenvectors.

In the spin-conserving limit, Eq.~(\ref{eq:block_hamiltonian}) yields a particularly simple result because $E_{s,+}+E_{s,-}=2\varepsilon_s$:
\begin{equation}
 M_z^{\rm orb}=\frac{e}{\hbar}\sum_{s,\eta}\int_{\rm BZ}\frac{d^2k}{(2\pi)^2}
 \Theta(\mu-E_{s,\eta})\,[\varepsilon_s(\mathbf k)-\mu] \,\Omega_{s,\eta}(\mathbf k).
 \label{eq:Morb_two_band}
\end{equation}
Hence the same Berry curvature that controls the anomalous Hall response also determines the equilibrium orbital moment, with an energy weighting relative to the chemical potential.  For an insulator at half filling, differentiation inside the bulk gap gives
\begin{equation}
 \frac{\partial M_z^{\rm orb}}{\partial\mu}=-\frac{e}{h}C=-\frac{\sigma_{xy}}{e},
 \qquad
 \widetilde M_z\equiv\frac{h}{e}M_z^{\rm orb}
 \ \Rightarrow\ 
 \frac{\partial\widetilde M_z}{\partial\mu}=-C.
 \label{eq:Mtilde_slope}
\end{equation}
The in-gap slope, rather than the absolute value of $M_z^{\rm orb}$, is therefore quantized by the Chern number.

\begin{figure*}[t]
    \centering
    \includegraphics[width=0.98\textwidth]{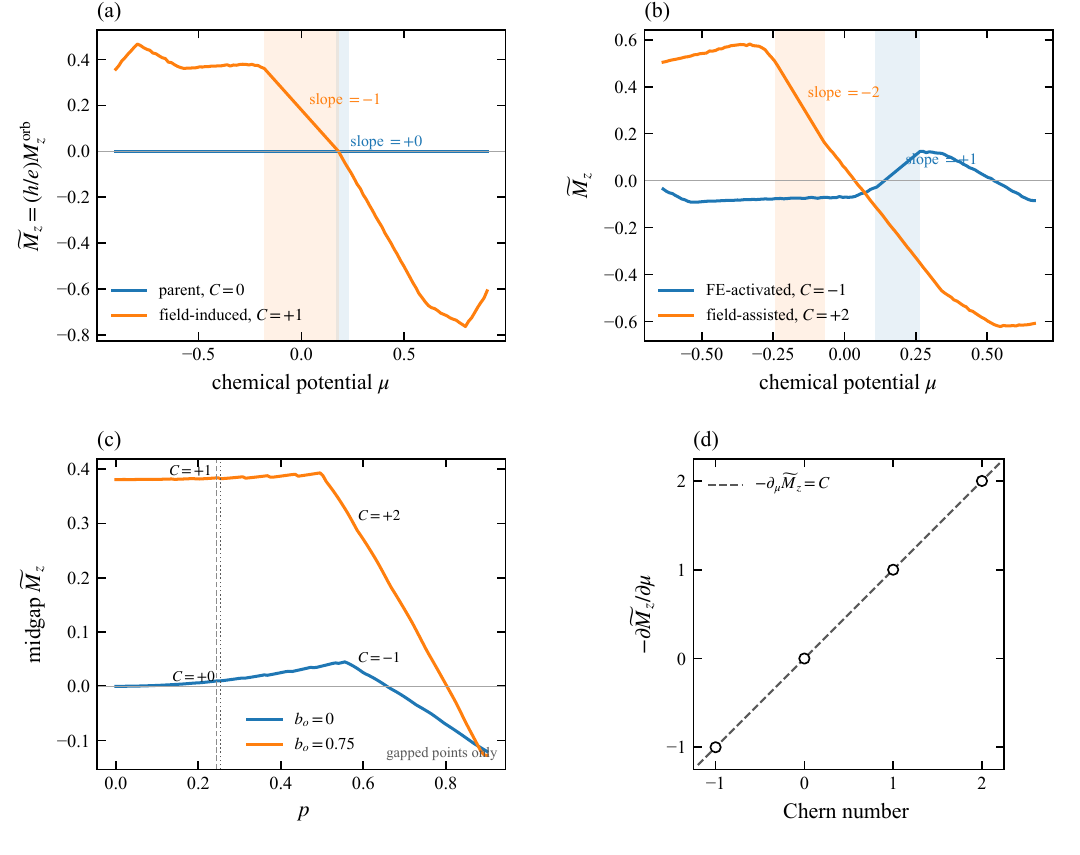}
    \caption{\label{fig:orbital_magnetization}
    Orbital magnetization and its topological relation to the Chern number.  (a) $\widetilde M_z=(h/e)M_z^{\rm orb}$ versus chemical potential for the compensated $C=0$ parent and field-induced $C=1$ phase.  (b) Corresponding curves for the ferroelectric $C=-1$ and field-assisted $C=2$ phases.  Shaded intervals mark global gaps; their slopes obey $\partial_\mu\widetilde M_z=-C$.  (c) Midgap orbital magnetization versus $p$ at $b_o=0$ and $0.75$; vertical lines mark the ferroelectric transitions.  (d) Numerical verification of $-\partial_\mu\widetilde M_z=C$ for representative $C=-1,0,1,2$ phases.  Common parameters are $A=1$, $B_0=0.40$, $M_0=1.20$, $\Delta_0=0.10$, and $\Delta_z=0.19$.}
\end{figure*}

Figure~\ref{fig:orbital_magnetization}(a) compares the compensated parent and the field-induced $C=1$ phase.  At $p=b_o=0$, $C_{4z}\mathcal T$ reverses $M_z^{\rm orb}$ while leaving the Hamiltonian invariant, so the orbital magnetization vanishes.  At $b_o=0.40$, the symmetry is broken and $\widetilde M_z$ varies linearly across the gap with slope $-1$, as required by $C=1$.  Figure~\ref{fig:orbital_magnetization}(b) shows the corresponding slopes $+1$ and $-2$ for the zero-field $C=-1$ and field-assisted $C=2$ phases.

The polarization dependence in Fig.~\ref{fig:orbital_magnetization}(c) separates symmetry breaking from topology.  At zero field, $M_z^{\rm orb}$ becomes finite as soon as $p\ne0$, even before the Chern transition at $p_c\simeq0.254$; for example, at $p=0.22$ the system still has $C=0$ but $\widetilde M_z\simeq7.9\times10^{-3}$. Thus, a finite orbital magnetization diagnoses the removal of the $C_{4z}\mathcal T$ cancellation, whereas the quantized slope in Eq.~(\ref{eq:Mtilde_slope}) diagnoses the Chern topology.  At $b_o=0.75$, the corresponding $C=1\to2$ transition occurs at $p_c\simeq0.244$ and is accompanied by a continuous but strongly polarization-dependent orbital response.  Figure~\ref{fig:orbital_magnetization}(d) confirms $-\partial_\mu\widetilde M_z=C$ for representative $C=-1,0,1,2$ states using the full four-band calculation.

\begin{figure*}[t]
    \centering
    \includegraphics[width=0.8\textwidth]{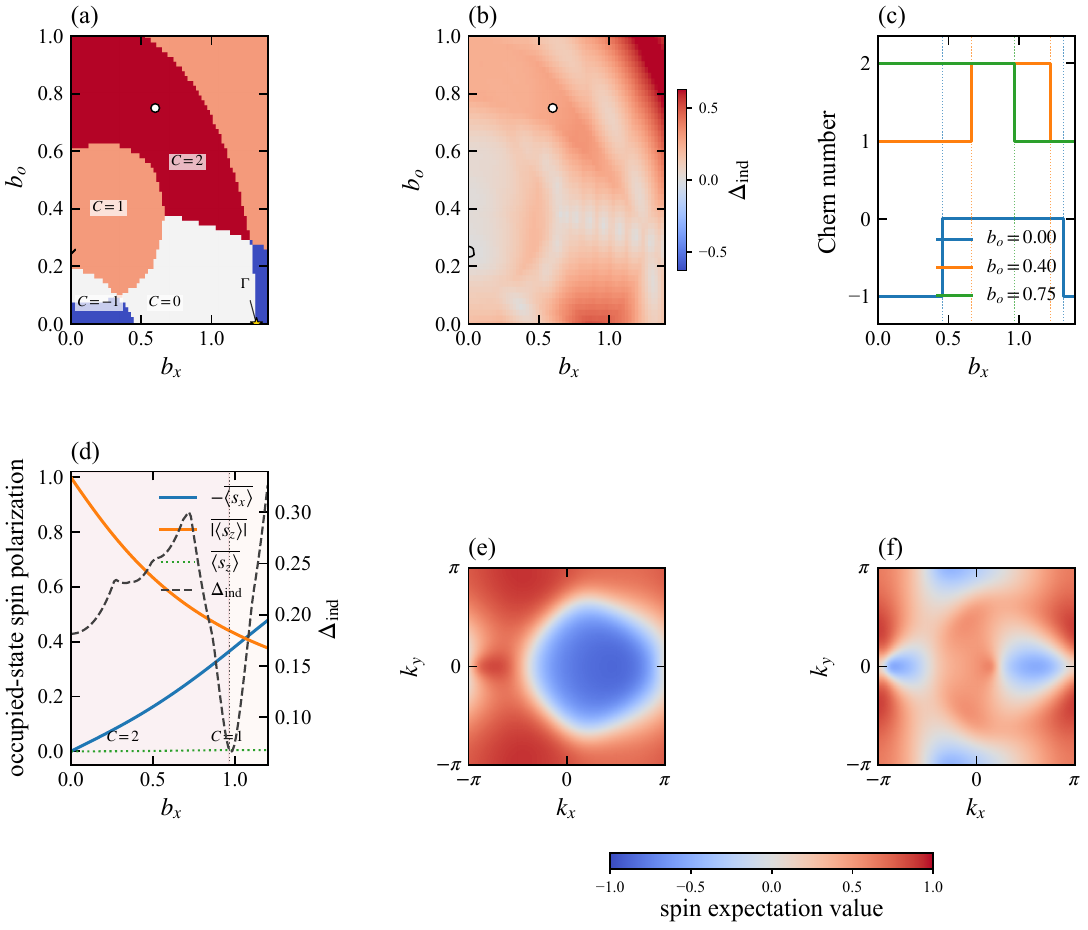}
    \caption{\label{fig:spin_mixing}
    Robustness to transverse magnetic mixing at $p=0.55$.  (a) Non-Abelian Chern number in the $(b_x,b_o)$ plane.  (b) Indirect gap on the same plane.  (c) Chern-number cuts for $b_o=0$, $0.40$, and $0.75$, showing trivialization, reentrance, and a canting-induced $C=1\to2$ transition.  (d) Brillouin-zone-averaged pseudospin observables and indirect gap for $b_o=0.75$.  (e),(f) Momentum-resolved $\langle s_z\rangle$ and $\langle s_x\rangle$ of the highest occupied band at $(p,b_o,b_x)=(0.55,0.75,0.60)$, where $C=2$.  Common parameters are $A=1$, $B_0=0.40$, $M_0=1.20$, $\Delta_0=0.10$, and $\Delta_z=0.19$.}
\end{figure*}

\subsection{Robustness to transverse spin mixing}
\label{subsec:spin_mixing}

The preceding analytical classification relies on $s_z$ conservation.  We now restore the transverse magnetic term $H_x=b_x\tau_0s_x$, for which $[H,s_z]\ne0$.  The sector invariants $C_\pm$ then cease to be meaningful, and the topology is characterized by the non-Abelian Chern number of the two occupied bands~\cite{fukui2005chern}.

A simple unitary relation constrains the phase diagram.  With $U_z=\tau_0s_z$, we have
 $U_zH(\mathbf k;b_x)U_z^\dagger=H(\mathbf k;-b_x)$, so $C(b_o,b_x,p)=C(b_o,-b_x,p)$,
while the transverse pseudospin polarization is odd in $b_x$.  It is therefore sufficient to consider $b_x\ge0$.

Additional analytical insight follows at $\Gamma$, where
\begin{align}
H_\Gamma&=M_0\tau_z+p\tau_x+b_o\tau_zs_z+b_xs_x,
 \\ \nonumber
 \det H_\Gamma&=(M_0^2+p^2-b_o^2-b_x^2)^2+4p^2b_o^2.
 \label{eq:gamma_det}
\end{align}
A $\Gamma$-point closing therefore requires $p b_o=0$ together with $M_0^2+p^2=b_o^2+b_x^2$.  When both $p$ and $b_o$ are finite, any Chern transition driven by $b_x$ must occur at finite momentum.

Figure~\ref{fig:spin_mixing}(a) shows the non-Abelian $C(b_x,b_o)$ map at $p=0.55$, while Fig.~\ref{fig:spin_mixing}(b) identifies the globally insulating regions.  The field cuts in Fig.~\ref{fig:spin_mixing}(c) reveal that transverse mixing can either suppress or enhance the Chern number.  At $b_o=0$, the $C=-1$ phase becomes trivial near $b_x\simeq0.45$ and reenters at the exact $\Gamma$-point condition $b_x=\sqrt{M_0^2+p^2}\simeq1.32$.  At $b_o=0.40$, the sequence is $C=1\to2\to1$, with transitions near $b_x\simeq0.65$ and $1.23$.  For $b_o=0.75$, the $C=2$ state persists to $b_x\simeq0.98$ before changing to $C=1$.  A projected two-band analysis of the finite-momentum closings confirms that each transition carries unit topological charge, consistent with the observed change in $C$ and with the Dirac-mass description of lattice Chern transitions~\cite{qi2006topological}.  

The spin mixing is substantial well before the topological gap closes.  At the representative point $(p,b_o,b_x)=(0.55,0.75,0.60)$, the non-Abelian invariant is $C=2$, with $\Delta_{\rm dir}\simeq0.37$ and $\Delta_{\rm ind}\simeq0.26$.  The Brillouin-zone-averaged transverse polarization is $\overline{\langle s_x\rangle}\simeq-0.20$, while the net longitudinal component remains nearly compensated, $\overline{\langle s_z\rangle}\simeq2\times10^{-3}$; the average magnitude $\overline{|\langle s_z\rangle|}\simeq0.58$ shows that a strong momentum-dependent longitudinal texture survives.  Figures~\ref{fig:spin_mixing}(e) and \ref{fig:spin_mixing}(f) display the corresponding $s_z$ and $s_x$ textures.  The quantized $C=2$ phase is therefore not an artifact of an exactly conserved spin sector.  Weak ferromagnetic responses generated by canting or anisotropic $g$ tensors have also been discussed in the context of altermagnets~\cite{jo2025weak,autieri2312staggered}.

\begin{figure*}[t]
\centering
\includegraphics[width=0.8\textwidth]{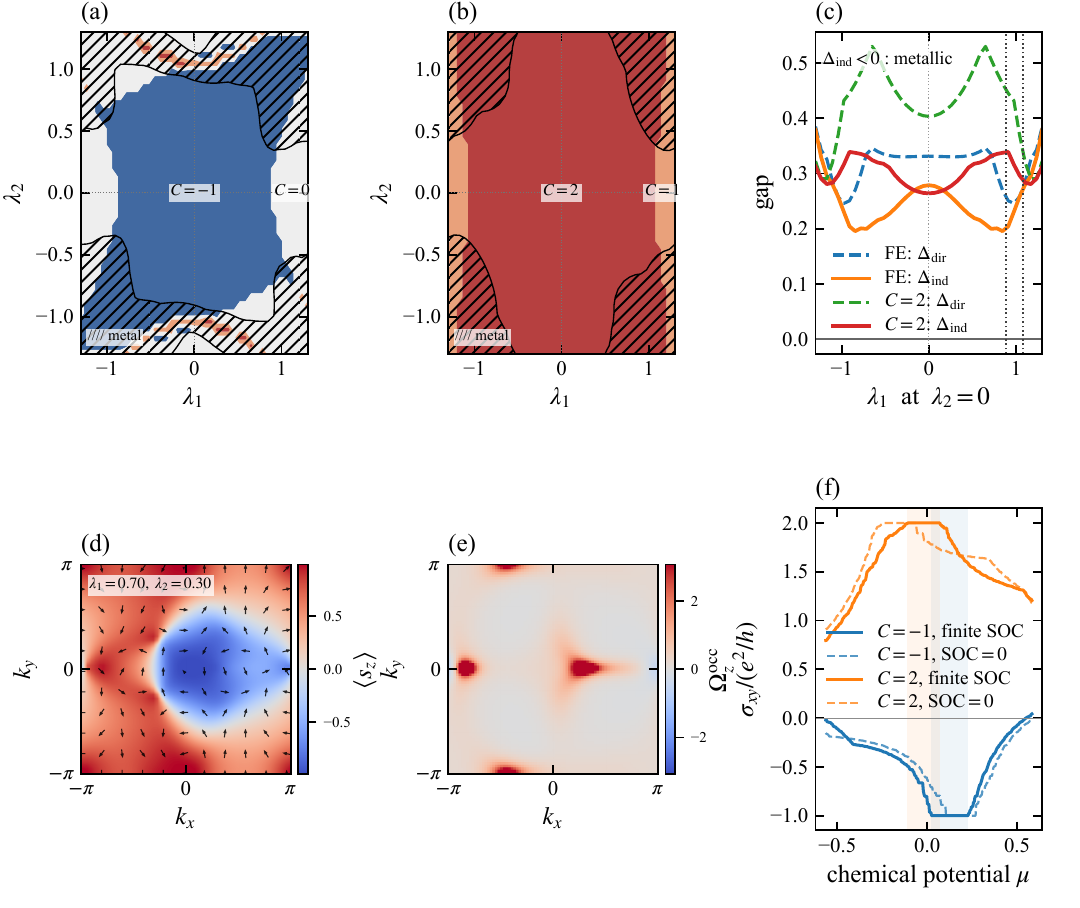}
\caption{\label{fig:SOCrobustness}
Robustness to symmetry-allowed SOC.  (a) Non-Abelian Chern number in the $(\lambda_1,\lambda_2)$ plane for the zero-field ferroelectric $C=-1$ state; hatching denotes a negative indirect gap.  (b) Corresponding map for the field-assisted $C=2$ state.  (c) Direct and indirect gaps along $\lambda_2=0$ for both states, showing that the zero-field phase becomes metallic before its direct topological transition, whereas the $C=2$ phase remains insulating to larger SOC.  (d) Momentum-resolved pseudospin texture of the highest occupied band at $(\lambda_1,\lambda_2)=(0.70,0.30)$ in the $C=2$ phase; arrows denote $(\langle s_x\rangle,\langle s_y\rangle)$ and color denotes $\langle s_z\rangle$.  (e) Occupied-subspace Berry curvature for the same state.  (f) Hall conductivity before (dashed) and after (solid) spin-mixing SOC for representative $C=-1$ and $C=2$ phases; shaded intervals mark the finite-SOC bulk gaps.  Common parameters are $A=1$, $B_0=0.40$, $M_0=1.20$, $\Delta_0=0.10$, and $\Delta_z=0.19$.}
\end{figure*}

\subsection{Robustness to symmetry-allowed spin--orbit coupling}
\label{subsec:soc_robustness}

Finally, we test whether the electrically controlled phases survive inversion-asymmetric SOC.  In the adopted double-group representation, the conventional orbital-independent Rashba form is not by itself $C_{4z}$ covariant.  The leading linear-in-momentum spin-mixing structures can instead be written as Eq.\eqref{eq:HR12}.
 If $\lambda_{1,2}$ originate from an inversion-odd structural asymmetry, inversion covariance is recovered by reversing $\lambda_{1,2}$ together with $\mathbf k$.  This construction is consistent with the appearance of inversion-asymmetric spin-off-diagonal terms in BHZ-type quantum wells~\cite{rothe2010fingerprint} and with electrically tunable Rashba-like coupling in polar materials~\cite{di2012electric}.

The in-plane polar order can also generate an odd-in-momentum SOC term diagonal in $s_z$; a representative $C_{4z}$-compatible invariant can be given as Eq.\ref{eq:HPSO}.
For the $x$-polarized domain, $H_{P\mathrm{SO}}\propto p\sin k_y$, while the ferroelectric Dirac closing in Eq.~(\ref{eq:FEkineticzeros}) satisfies $\sin k_y^c=0$.  Hence $H_{P\mathrm{SO}}(\mathbf k_c)=0$: this channel leaves the direct critical polarization unchanged ($p_c\simeq0.254$ for the parameters used here), although it can modify the indirect gap and the Berry-curvature distribution away from the critical momentum.  Without additional mirror constraints, Eq.~(\ref{eq:HPSO}) should be viewed as a representative, rather than unique, polar SOC invariant.

For finite $\lambda_{1,2}$, $s_z$ is not conserved and we again use the non-Abelian Chern number.  Figure~\ref{fig:SOCrobustness}(a) shows that the zero-field $C=-1$ phase at $p=0.55$ survives over a broad region of the $(\lambda_1,\lambda_2)$ plane.  Along $\lambda_2=0$, the indirect gap closes at $|\lambda_1|\simeq0.76$, whereas the direct topological transition occurs only near $\lambda_1\simeq0.88$ [Fig.~\ref{fig:SOCrobustness}(c)].  The intermediate regime is therefore metallic at half filling even though the directly separated occupied subspace still carries $C=-1$.

The field-assisted $C=2$ phase is more robust.  For $p=0.55$ and $b_o=0.75$, a broad $C=2$ region surrounds the spin-conserving point [Fig.~\ref{fig:SOCrobustness}(b)], and along $\lambda_2=0$ the transition to $C=1$ occurs only near $\lambda_1\simeq1.08$.  The state therefore remains topological after the two Kramers sectors are strongly hybridized.

Figure~\ref{fig:SOCrobustness}(d) displays the momentum-resolved
pseudospin texture of the highest occupied band for the representative
SOC-mixed state
$(p,b_o,\lambda_1,\lambda_2)=(0.55,0.75,0.70,0.30)$.
The arrows represent the in-plane components
$(\langle s_x\rangle,\langle s_y\rangle)$, while the color scale denotes
the out-of-plane component $\langle s_z\rangle$.
Despite the pronounced noncollinear texture, the occupied subspace retains
the non-Abelian Chern number $C=2$ and an indirect gap of approximately
$0.18$. The mean magnitude of the in-plane pseudospin is about $0.51$,
whereas $\overline{|\langle s_z\rangle|}\simeq0.58$.
Thus, the $C=2$ phase remains robust even when the Kramers pseudospin is
strongly rotated away from the $s_z$ axis, demonstrating that its topology
does not rely on approximate spin conservation.
 
 SOC also redistributes the occupied Berry curvature into finite-momentum hot spots [Fig.~\ref{fig:SOCrobustness}(e)] without changing its integrated value.  The Hall curves in Fig.~\ref{fig:SOCrobustness}(f) provide the corresponding transport check: the finite-SOC $C=-1$ and $C=2$ states retain plateaus at $-e^2/h$ and $2e^2/h$, respectively, while the nonquantized response outside the gaps is substantially reconstructed.  In this sense, the FE/field terms create the Chern phase, whereas SOC reshapes the wave functions and geometric response without generically destroying the quantized topology.  This distinction is consistent with the nonrelativistic origin of the altermagnetic splitting~\cite{vsmejkal2022giant} and with the known sensitivity of Berry-curvature-driven Hall responses to SOC in compensated magnets~\cite{doi:10.1126/sciadv.aaz8809}.

\begin{figure*}[t]
\centering
\includegraphics[width=0.8\textwidth]{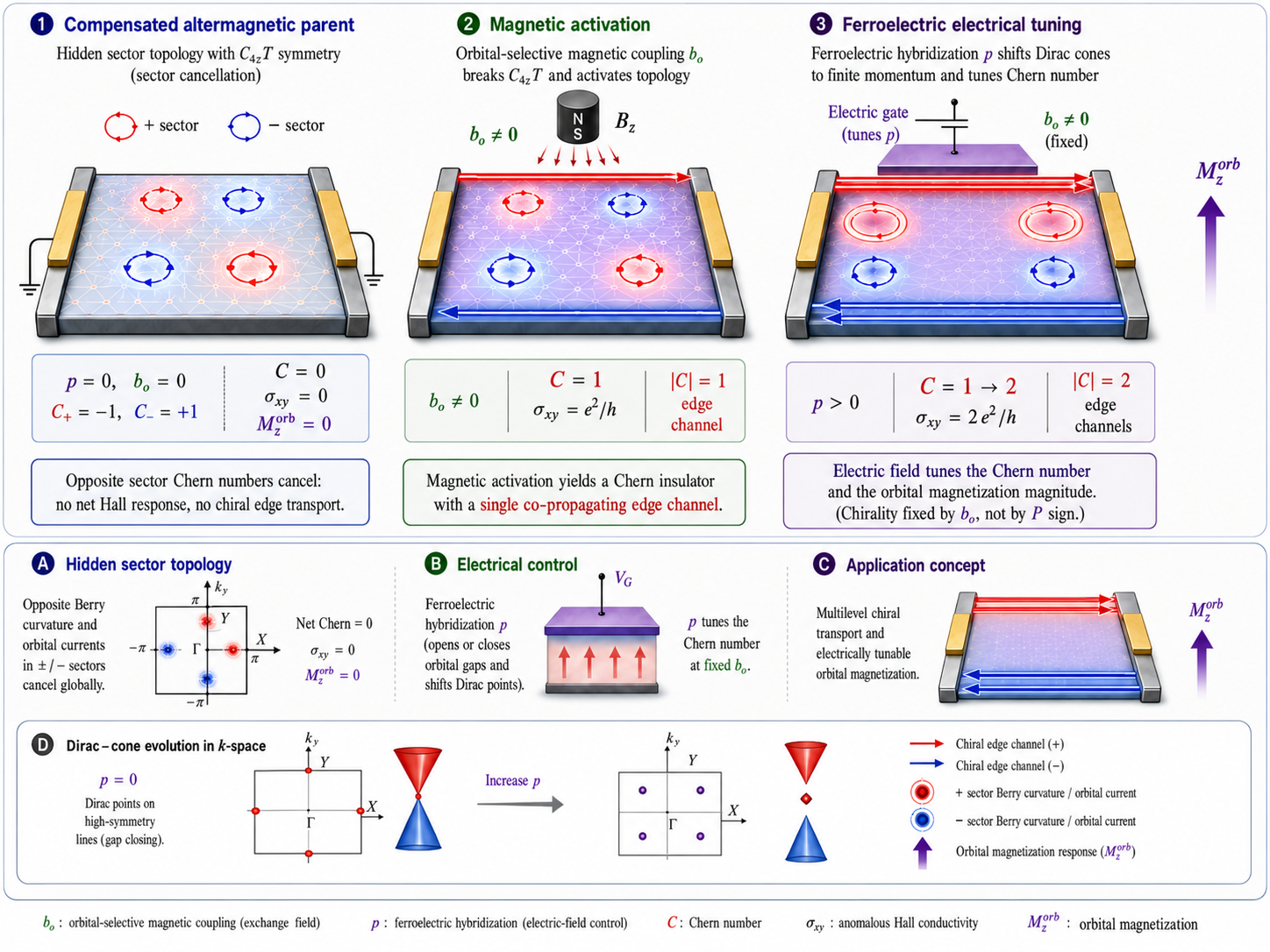}
\caption{\label{fig:summary_schematic}
\textbf{Conceptual summary of magnetic activation and ferroelectric control of Chern topology in a compensated altermagnet.}
(1) In the nonpolar, zero-field reference state, the two altermagnetic sectors carry opposite topological character, $C_{+}=-1$ and $C_{-}=+1$. The combined $C_{4z}\mathcal T$ symmetry enforces cancellation of their Berry-curvature contributions, giving $C=0$, $\sigma_{xy}=0$, and vanishing orbital magnetization. The red and blue circulating motifs schematically represent opposite sector-resolved Berry-curvature/orbital-current contributions rather than literal real-space current loops.
(2) An orbital-selective magnetic coupling $b_o$ breaks the compensating symmetry and unbalances the two sectors, producing a Chern-insulating state with $C=1$, a quantized Hall conductivity $\sigma_{xy}=e^2/h$, and one chiral edge channel per boundary.
(3) At fixed magnetic coupling, the ferroelectric orbital hybridization $p$ provides an electrical control parameter that displaces the Dirac closing to finite momentum and can tune the system from $C=1$ to $C=2$. The latter phase supports two co-propagating chiral channels on each edge, $\sigma_{xy}=2e^2/h$, and a finite, electrically tunable orbital magnetization. In the minimal model, the chirality is controlled by the magnetic sector imbalance rather than by the sign of the ferroelectric polarization alone.
The lower panels summarize the hidden sector topology, gate-controlled orbital hybridization, the resulting multilevel chiral-transport concept, and the finite-momentum displacement of the Dirac criticality. The schematic illustrates the operating principle of the proposed ferroelectric altermagnetic Chern-insulator platform and is not intended to represent a specific microscopic material geometry.
}
\end{figure*}
 
\section{Discussion and conclusions}
\label{sec:discussion}

The results above establish a simple mechanism for converting a compensated
altermagnetic band structure into an electrically tunable Chern insulator.
The essential starting point is not a topologically trivial collection of
spin-degenerate bands, but a band-inverted altermagnetic state in which the
two Kramers sectors carry opposite topological character.  In the
$C_{4z}\mathcal T$-symmetric reference state these contributions cancel
exactly, giving a vanishing charge Chern number despite pronounced
momentum-dependent spin splitting and sector-resolved Berry curvature.  The
central role of ferroelectricity and magnetic coupling is therefore to remove
this symmetry-enforced cancellation selectively.  The orbital-dependent
magnetic term changes the Dirac masses of the two sectors unequally, whereas
the polar orbital hybridization displaces the Dirac points away from the
time-reversal-invariant momenta.  Their combined action provides access to
$C=0$, $\pm1$, and $\pm2$ insulating phases without requiring a large uniform
ferromagnetic moment.

The orbital-selective magnetic coupling used here has a close analogue in
magnetized BHZ systems.  In HgMnTe quantum wells, the different exchange
responses of the electron- and hole-like orbital sectors were shown to be
crucial for converting the quantum spin Hall state into a quantum anomalous
Hall state~\cite{liu2008quantum}.  In the present model, $b_o\tau_zs_z$ plays the
corresponding orbital-difference role, but acts on an altermagnetic background
whose nonrelativistic spin splitting is already momentum-dependent.  This
distinction allows odd- and even-Chern phases to emerge through an imbalance
of initially compensating altermagnetic sectors.  We have treated $b_o$ as a
phenomenological magnetic energy scale controlled by the longitudinal
magnetic environment.  Since orbital Landau quantization has intentionally
been neglected, the present description applies most directly when
Zeeman/exchange effects dominate the relevant low-energy reconstruction, or
when an effective orbital-dependent exchange field is generated without a
large orbital magnetic response.  A material-specific treatment at sizable
perpendicular magnetic field would require inclusion of Peierls phases and
Landau-level formation in addition to the couplings considered here.

Ferroelectricity provides a qualitatively different control parameter because
it acts directly on the orbital hybridization rather than on the magnetic
moment.  This mechanism is consistent with the rapidly developing concept of
ferroelectric altermagnetism, in which electric polarization can be coupled
to momentum-dependent altermagnetic spin splitting without requiring reversal
of the N\'eel order~\cite{gu2025ferroelectric}.  Importantly, the electrical
control predicted by our minimal Hamiltonian should not be confused with an
automatic reversal of Hall chirality upon switching between inversion-related
ferroelectric domains.  Because the phase boundaries depend on the
$x$-polarized hybridization through its magnitude, the minimal model obeys
$C(+p,b_o)=C(-p,b_o)$.  Electrical switching therefore refers here to changing
the magnitude of the polar distortion, or to switching between nonpolar and
polar structures with different Chern numbers.  Reversing the Chern number
solely by $P\rightarrow-P$ would require an additional symmetry-breaking
coupling that makes the two polar domains electronically inequivalent.  This
qualification is also consistent with symmetry analyses of ferroelectric
altermagnets, which show that coexistence of ferroelectricity and
altermagnetism alone does not guarantee reversal of the altermagnetic
electronic structure under polarization switching~\cite{gu2025ferroelectric}.

The required ingredients are not restricted to a single microscopic material
class.  First-principles studies already identify low-dimensional systems in
which tetragonal altermagnetism, sizable momentum-dependent spin splitting,
band inversion, and polar order coexist.  For example, Janus
Cr$_2$SeO has been predicted to exhibit a compensated altermagnetic electronic
structure with opposite spin characters near the $X$ and $Y$ valleys,
substantial spin splitting, band inversion, and finite ferroelectric
polarization~\cite{khan2025altermagnetism}.  Its reported polarization is predominantly out-of-plane, so it should not be regarded as a direct realization of the in-plane polar Hamiltonian studied here.  Rather, it demonstrates that several of the ingredients entering the minimal model can coexist in a single
two-dimensional tetragonal material.  Together with the broader class of
ferroelectrically switchable altermagnets identified by symmetry and
first-principles screening~\cite{gu2025ferroelectric}, this suggests two natural
materials routes: intrinsic multiferroic altermagnets and heterostructures in
which a band-inverted altermagnet is coupled to a ferroelectric layer or polar
interface.  Establishing a specific realization will require constructing a
Wannier Hamiltonian for the candidate system and identifying the microscopic
counterparts of $p$, $b_o$, and the orbital-dependent altermagnetic exchange
$\Delta_z$.

Several experimental signatures distinguish the mechanism proposed here.
The most direct transport signature is the quantized Hall response
$\sigma_{xy}=C e^2/h$, accompanied by $|C|$ co-propagating chiral modes on
each edge.  In particular, observation of two edge channels and a
$2e^2/h$ plateau would directly identify the field-assisted $C=2$ state.
The ferroelectric transition also has a characteristic momentum-space
signature: the bulk gap closes away from the conventional high-symmetry
point because the polar hybridization shifts the zero of the Dirac kinetic
term.  Momentum-resolved spectroscopy should therefore search for the
closing and reopening of the gap at a displaced momentum rather than only
along the endpoints of the conventional high-symmetry path.  Such
spectroscopic characterization is realistic in the broader altermagnetic
context: broken Kramers degeneracy and large nonrelativistic spin splitting
have already been resolved by angle-resolved photoemission in
$\alpha$-MnTe~\cite{lee2024broken}.  Moreover, nanoscale altermagnetic domains
in MnTe have been imaged using complementary x-ray dichroism techniques and
their configurations manipulated by microstructuring and magnetic-field
cooling~\cite{Amin2024MnTe}.  These advances demonstrate that both the
momentum-space electronic structure and the real-space magnetic state of
altermagnets are experimentally accessible.

The orbital magnetization provides a complementary bulk diagnostic.  Its
absolute value is not quantized and can become finite whenever
$C_{4z}\mathcal T$ is broken, including on the topologically trivial side of
a transition.  By contrast, within an insulating gap the chemical-potential
slope satisfies $\partial_\mu\widetilde M_z=-C$.  The same integer that fixes
the edge-state multiplicity and the Hall plateau therefore also fixes a
thermodynamic response.  This distinction is useful experimentally because a
finite orbital moment alone does not establish Chern topology, whereas its
gate-dependent in-gap slope provides an independent test of the topological
index.  The simultaneous observation of a quantized Hall plateau, the
corresponding number of chiral edge channels, and the predicted orbital
magnetization slope would provide a stringent identification of the phase.

The robustness calculations further show that the mechanism is not an
artifact of exact $s_z$ conservation.  A transverse Kramers-sector mixing
term can generate strongly noncollinear pseudospin textures while preserving
the bulk Chern number over a substantial parameter range, and can even drive
transitions between $C=1$ and $C=2$ phases.  Such transverse mixing should
be viewed phenomenologically rather than identified universally with a
specific microscopic canting mechanism.  Nevertheless, weak ferromagnetic
responses and transverse magnetic components are symmetry allowed in subsets
of altermagnets, and alternating $g$-tensor anisotropy has recently been
identified as one microscopic route to weak ferromagnetism even in cases
where conventional Dzyaloshinskii--Moriya canting is absent
~\cite{jo2025weak}.  Likewise, the symmetry-derived inversion-asymmetric
SOC terms considered here strongly reconstruct the pseudospin and Berry
curvature distributions but leave the integer Hall response unchanged until
the bulk gap closes.  The Chern phases are therefore properties of the full
occupied-band manifold rather than of an artificial decomposition into
conserved spin blocks.

The present model is intentionally minimal.  It does not include disorder,
electron correlations beyond the effective single-particle parameters,
finite-temperature fluctuations of the altermagnetic or ferroelectric order,
or orbital coupling to an external magnetic field.  These effects will
determine the quantitative gap sizes and experimental operating regime of a
specific compound.  They do not, however, alter the symmetry-based mechanism:
an integer Chern number cannot change under a perturbation that preserves the
global insulating gap.  The most important materials requirements are
therefore a band-inverted two-orbital manifold near the Fermi energy, a
compensated altermagnetic order producing symmetry-related spin-split
sectors, a controllable polar distortion that couples efficiently to the
orbital hybridization, and a magnetic or exchange environment capable of
producing an orbital-dependent sector imbalance.

The operating principle emerging from these results is summarized schematically in Fig.~\ref{fig:summary_schematic}. In the compensated reference state, the two altermagnetic sectors possess opposite topological character, but their contributions cancel because of the $C_{4z}\mathcal T$ symmetry, leaving no net anomalous Hall response. An orbital-selective magnetic coupling removes this cancellation and converts the latent sector topology into a finite charge Chern number. Ferroelectric orbital hybridization then provides an independent electrical control parameter: by shifting the Dirac criticality to finite momentum, it can change the number of topological channels at fixed magnetic coupling, for example from $C=1$ to $C=2$. The resulting change is directly reflected in the quantized Hall conductivity, the number of chiral edge modes, and the orbital-magnetization response. Importantly, in the minimal model the electric field controls the topological sector through the magnitude of the polar hybridization, whereas the sign of the Hall chirality remains tied to the magnetic sector imbalance rather than to the sign of the polarization itself. Figure~\ref{fig:summary_schematic} therefore emphasizes the central design principle of the present work: magnetic symmetry breaking activates the uncompensated Chern response, while ferroelectricity provides electrical control over its topological multiplicity.

In summary, we have shown that a compensated $d$-wave altermagnet can host a
latent topological structure in which opposite sector Chern numbers cancel
in the high-symmetry reference state.  Magnetic orbital selectivity removes
this cancellation and generates $C=\pm1$ and $\pm2$ phases, while
ferroelectric orbital hybridization provides an independent electrical knob
that moves the topological Dirac points to finite momentum and changes the
number of chiral channels.  Bulk Chern numbers, ribbon spectra, quantized
Hall transport, and orbital magnetization give mutually consistent
signatures of these phases.  Their persistence under transverse
Kramers-sector mixing and symmetry-allowed spin--orbit coupling demonstrates
that the mechanism does not rely on exact spin conservation.  More broadly,
the results identify \emph{symmetry-selective activation of compensating
topological sectors} as a route to electrically controllable Chern phases in
altermagnetic systems, and provide a minimal framework for searching for
ferroelectric altermagnetic Chern insulators in intrinsic multiferroics and
engineered two-dimensional heterostructures.

\vspace{5mm}
\begin{acknowledgments}
M.B.T. acknowledges the funding support by  Narodowa Agencja Wymiany Akademickiej (NAWA) under the ULAM program with project number BPN/ULM/2025/1/00156/U/00001.
M.B.T. acknowledges the funding support by Iran National Science Foundation (INSF) under project No.4043973. This research was supported by the Foundation for Polish Science project “MagTop” no. FENG.02.01-IP.05-0028/23 co-financed by the European Union from the funds of Priority 2 of the European Funds for a Smart Economy Program 2021–2027 (FENG). We acknowledge the access to the computing facilities of the Poznan Supercomputing and Networking Center, Grant No. pl0807.
\end{acknowledgments}

\medskip

\appendix

\bibliography{references}
\end{document}